\documentclass[pra,nobibnotes,aps,showpieces,superscriptaddress,amsfonts,amsmath,float fix,backend=bibtex]{revtex4}
\usepackage[capitalize]{cleveref}
\usepackage{graphicx}
\usepackage{color}
\usepackage{import}
\usepackage{wrapfig}
\usepackage{mathtools}
\usepackage{bm}
\usepackage{amsmath}
\usepackage{amsfonts}
\usepackage{amssymb}

\begin{document}


\title{The amplitude decay of a harmonic oscillator damped simultaneously by weak linear and nonlinear damping forces}

\def\correspondingauthor{\footnote{Corresponding author}}

\author{Karlo Lelas\correspondingauthor{}}
\email{klelas@ttf.unizg.hr}
\affiliation{Faculty of Textile Technology, University of Zagreb, Prilaz baruna Filipovića 28a, 10000 Zagreb, Croatia}

\author{Robert Pezer}
\email{rpezer@simet.unizg.hr}
\affiliation{Department of Physical Metallurgy, Faculty of Metallurgy, University of Zagreb, Aleja narodnih heroja 3, 44000 Sisak, Croatia}

\date{\today}

\begin{abstract}
We derive approximate expressions for the amplitude decay of harmonic oscillations weakly damped by the simultaneous action of three different damping forces: force of constant magnitude, force linear in velocity, and force quadratic in velocity. Our derivation is based on a basic understanding of the undamped harmonic oscillator and the connection between the energy dissipation rate and the power of the total damping force. By comparing our approximate analytical solutions with the corresponding numerical solutions, we find that our solutions excellently describe the dynamics of the oscillator in the regime of weak damping by combinations of these three forces for an experimentally relevant range of corresponding damping constants. The physical concepts and mathematical techniques we employ are suitable for undergraduate physics teaching.
\end{abstract}

\maketitle

\section{Introduction}

In undergraduate physics textbooks, see, e.g., \cite{Cutnell8, Resnick10, Young2020university, Waves}, the damping of harmonic oscillations is considered only in the case of viscous damping, i.e. for a damping force linear in velocity. The damping force of constant magnitude and the damping force quadratic in velocity are covered only in the context of unidirectional motions such as sliding on an inclined plane with friction \cite{Cutnell8, Resnick10, Young2020university} and the motion of a free falling object with air resistance \cite{Resnick10, Young2020university}, while the analysis of the influence of these forces on the dynamics of the harmonic oscillator is omitted. The reason for this is that these two damping forces require a mathematical treatment that is demanding for undergraduates, while on the other hand the dynamics of a harmonic oscillator damped by a force linear in velocity can be analyzed in a fashion known to students, i.e. the solution of the corresponding equation of motion can be written analytically in a closed form valid for all times, and such a solution is easy to analyze. In case of a harmonic oscillator damped with sliding friction (often called Coulomb damping), the corresponding equation of motion can be solved exactly by splitting the motion into left and right moving segments \cite{Lapidus, AviAJP, Grk2, Kamela}, i.e. the motion needs to be analyzed over half-cycles and the solution thus obtained cannot be put in a closed form valid for all times. Furthermore, in the case of sliding friction, the oscillations generally do not halt at the equilibrium position, which cannot be directly read from the exact solution of the equation of motion, but it is necessary to additionally analyze the instant of the halt \cite{Lapidus, AviAJP, Grk2, Kamela}. In the case of a harmonic oscillator damped with air resistance, the corresponding equation of motion is a nonlinear differential equation that cannot be solved analytically and approximate or numerical methods must be used \cite{Smith, Mungan, Wang, Grk2}. 

Even in the simplest real-world systems, damping of vibrations is caused by combinations of two or all three of the above forces, with each of them contributing to damping to a greater or lesser extent depending on the specifics of the system. For example, in \cite{AJPpendulum} a simple physical pendulum was studied, both experimentally and theoretically, and it was shown that the contribution of all three forces, i.e. sliding friction in the pendulum support and air resistance with both linear and quadratic terms in velocity, must be taken into account for an adequate general description of pendulum damping, while in some specific cases two (out of three) damping forces were sufficient. The damping of harmonic oscillations by the combined action of sliding friction and viscous damping, as well as the combination of viscous and quadratic damping, has been investigated by many authors. 
For example, the amplitude decay in the case of free vibrations with combined Coulomb and viscous damping was first derived under no-stop conditions in \cite{Markho}, while the theory of amplitude decay presented in \cite{AJPRicch} takes into account the stopping of dynamics with an offset from the equilibrium position. The exact theory of the discontinuities in the acceleration of such systems is given and experimentally verified in \cite{AJPHinrich}. 
In the case of combined linear and quadratic damping, the approximate analytical expression for amplitude decay was derived using the work energy theorem, i.e. the energy dissipated due to damping over the first half period, already in \cite{AJPNelson}, while numerical methods are routinely used, see, e.g., \cite{Bacon}. 

Similarly as in \cite{AJPNelson}, the connection of the energy dissipation rate and the power of the damping force was used to determine the amplitude decay in case of weak damping by each of the three damping forces individually in \cite{Wang} and by the combination of sliding friction and viscous damping in \cite{Vitorino}.
More precisely, in the case of weak damping, one can take that the amplitude remains approximately constant over half cycles (half periods) and average the oscillatory parts of the energy dissipation rate over these time intervals to obtain the first-order differential equation for the time-dependent amplitude \cite{Wang}, which is easy to solve using the separation of variables method. The validity of the thus obtained approximate analytical solutions was confirmed by experiments \cite{AJPNelson, Wang, Vitorino}. Although it is straightforward, to the best of our knowledge, this approach has not been used for the analysis of the harmonic oscillator weakly damped by a combination of all three damping forces in the literature so far.         

Here we use the approach presented in \cite{Wang} to obtain approximate analytical expressions for the amplitude decay of harmonic oscillations weakly damped by a combination of three different damping forces: force of constant magnitude, force linear in velocity, and force quadratic in velocity. By comparing our approximate analytical solutions with the corresponding numerical solutions, we show that our solutions excellently describe the dynamics of the oscillator in the regime of weak damping by simultaneous action of these three forces. In addition, we have not been able to find examples in the literature where this approach is used to consider damping of oscillations by a combination of sliding friction and force quadratic in velocity. Therefore, to cover this case and for completeness, we also analyze the amplitude decay in cases of weak damping by a combination of any two of the three considered damping forces. On the mathematical side, our derivation is based on basic knowledge of integral calculus, and our approximate analytical solutions are expressed by elementary functions, i.e. both are well known to undergraduate students. Thus, the approach and results presented here are suitable for undergraduate physics classes.

This paper is organized into five sections. In Section \ref{Basic}, we present the theory and approximations we use and derive the first-order differential equation for the amplitude decay. In Section \ref{all3}, we obtain solutions in the case of damping by a combination of three damping forces. In Section \ref{comb2}, we obtain solutions in the case of damping by a combination of any two of the three considered damping forces. In Section \ref{dodatak}, we analyze limitations of our approximate analytical solutions and comment on possible improvements to the presented approach. In Section \ref{Conclusion}, we briefly comment on experimental setups in which our theoretical results could be tested and summarize important findings of the paper.

\section{Basic theory, approximations we use, and equation of the amplitude decay}
\label{Basic}

\begin{figure}[h!t!]
\begin{center}
\includegraphics[width=0.65\textwidth]{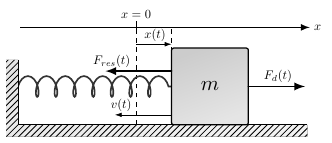}
\end{center}
\caption{Schematic representation of a block-spring system with a restoring force $F_{res}(t)$ and a (total) damping force $F_d(t)$. Here we show the time instant at which $x(t)>0$ and $v(t)<0$, i.e. at which $F_{res}(t)<0$ and $F_d(t)>0$. See text for details.} 
\label{shema}
\end{figure}

As an example of a damped harmonic oscillator, we consider the system schematically shown in Fig.\,\ref{shema}, i.e. a block of mass $m$ that oscillates back and forth along a horizontal surface under the influence of the restoring force of an ideal spring $F_{res}(t)=-kx(t)$, where $k$ is the stiffness of the spring and $x(t)$ is the displacement of the block from the equilibrium position (set to $x=0$), and the damping force
\begin{equation}
F_d(t)=-\text{sgn}[v(t)]\mu mg-bv(t)-Dv(t)|v(t)|\,,
\label{Fd}
\end{equation}
where $\mu>0$, $b>0$ and $D>0$ are the corresponding damping constants, $v(t)=dx(t)/dt$ is the velocity, and  
\begin{equation} \label{signum}
    \text{sgn}\left[v(t)\right] = \begin{cases}
\begin{tabular}{@{}cl@{}}
   $1$\, & if\, $v(t)$ $ > $ $0$ \\
    $0$\, & if\, $v(t)$ $=$ $0$ \\
    $-1$\, & if\, $v(t)$ $<$ $0$\,
\end{tabular}
    \end{cases}
    \end{equation}
is the sign function. The corresponding equation of motion is
\begin{equation}
ma(t)=-\text{sgn}\left[v(t)\right]\mu mg-bv(t)-Dv(t)|v(t)|-kx(t)\,,
\label{HOeq}
\end{equation}
where $a(t)=d^2x(t)/dt^2$ is the acceleration of the block. The first term on the right-hand side of equation \eqref{HOeq} models the force of sliding friction \cite{Lapidus, AviAJP, Grk2, AJPHinrich}, while the second and third terms model, e.g., the influence of air resistance, since air resistance generally depends on terms proportional to velocity and square of velocity \cite{AJPpendulum, AJPNelson, Bacon, Wang}. In Section \ref{Conclusion} we comment in more detail on physical systems where combinations of all three types of damping occur. The energy (potential plus kinetic) of the block-spring system is given by  
\begin{equation}
E(t)=\frac{kx^2(t)}{2}+\frac{mv^2(t)}{2}\,.
\label{Energy}
\end{equation}
If we put $\mu=b=D=0$ in \eqref{HOeq}, i.e. for $F_d(t)=0$, we get the equation of the undamped harmonic oscillator \cite{Resnick10}, with general solution 
\begin{equation}
x_{HO}(t)=A_0\cos(\omega_0t+\varphi_0)\,,
\label{xHO}
\end{equation}
where $\omega_0=\sqrt{k/m}$ is the angular frequency of the undamped system, while $A_0$ and $\varphi_0$ are constants (amplitude and initial phase) that are determined from initial conditions $\left(x_0,v_0\right)$, i.e. $A_0=\sqrt{x_0^2+(v_0/\omega_0)^2}$ and $\varphi_0=\arctan\left(-v_0/(\omega_0x_0)\right)$.
The undamped system oscillates with conserved, i.e. constant, energy \cite{Resnick10}. For damped systems, i.e. for $F_d(t)\neq0$, the energy is not conserved due to the power of the damping force $P_d(t)=F_d(t)v(t)$, and the energy dissipation rate is given by \cite{Young2020university} 
\begin{equation}
\frac{dE(t)}{dt}=F_d(t)v(t)=-\mu mg|v(t)|-bv^2(t)-D|v(t)|^3\,,
\label{Power}
\end{equation}
where we used $\text{sgn}\left[v(t)\right]v(t)=|v(t)|$ and $v^2(t)|v(t)|=|v(t)|^3$. Thus, equation \eqref{Power} tells us that the energy of the damped system decreases monotonically with time, i.e. $\frac{dE(t)}{dt}\leq0$ for all $t\geq0$, and $\frac{dE(t)}{dt}=0$ holds at the turning points, i.e. at instants when $v(t)=0$.

All three types of damping are analyzed both theoretically and experimentally in \cite{Wang}. Considering, e.g., the graphical representations of the experimental results presented in \cite{Wang}, we can easily conclude that the defining characteristics of the weak damping regime, for each individual type of damping, are that the frequency and the initial phase remain approximately the same as in the undamped case, while the amplitude of the oscillations decreases very slowly over time, i.e. it decreases only slightly during one full oscillation (one period) of the system. Of course, such a conclusion can also be supported by purely theoretical considerations of the weak damping limit of exact analytical solutions in the case of viscous damping or sliding friction, while quadratic damping requires a more complex analytical or a simple numerical analysis, as we show in Section \ref{dodatak}. Thus, in the case of weak damping with all three damping forces, it is reasonable to assume that the solution of equation \eqref{HOeq} is approximately of the form
\begin{equation}
x(t)=A_0f(t)\cos(\omega_0t+\varphi_0)\,,
\label{Xansatz1}
\end{equation}
where $f(t)$ is the unknown function that describes the amplitude decrease over time. The corresponding velocity is
\begin{equation}
v(t)=\frac{dx(t)}{dt}=-\omega_0A_0\left(f(t)\sin(\omega_0t+\varphi_0)-\frac{df(t)}{dt}\omega_0^{-1}\cos(\omega_0t+\varphi_0)\right)\,.
\label{Vansatz1}
\end{equation}
%
Since, by assumption, the amplitude changes very slowly compared to the oscillating part of \eqref{Xansatz1}, the rate of change of the function $f(t)$ over time has to be much slower than the rate of change of $\cos(\omega_0t+\varphi_0)$, which is of the order $\omega_0$. We can conclude that $|df(t)/dt|\ll\omega_0$ holds for weak damping for all time, i.e., for $t\geq0$ (we have put absolute value because we can expect the negative derivative of $f(t)$ since the amplitude decreases with time). Thus, due to $|df(t)/dt|\,\omega_0^{-1}\ll1$, we can take that the velocity is approximately of the form 
\begin{equation}
v(t)=-\omega_0A_0f(t)\sin(\omega_0t+\varphi_0)\,
\label{Vansatz11}
\end{equation}
It is easy to see that $f(0)=1$ must hold for the functions \eqref{Xansatz1} and \eqref{Vansatz11} to satisfy the initial conditions $x(0)=x_0$ and $v(0)=v_0$. Since $f(0)=1$ holds and $f(t)$ monotonically decreases towards zero, we can easily conclude that $f(t)\geq0$ must hold for all $t\geq0$. 
Using displacement \eqref{Xansatz1} and velocity \eqref{Vansatz11} in \eqref{Energy}, we get the corresponding approximate energy
\begin{equation}
E(t)=\frac{m\omega_0^2A_0^2f^2(t)}{2}\,.
\label{Energy1}
\end{equation}
The first derivative of \eqref{Energy1} is
\begin{equation}
\frac{dE(t)}{dt}=m\omega_0^2A_0^2f(t)\frac{df(t)}{dt}\,.
\label{derivEnergy12}
\end{equation}
Using \eqref{derivEnergy12} to approximate the left hand side of the energy dissipation rate \eqref{Power}, and velocity \eqref{Vansatz11} to approximate the right hand side of \eqref{Power}, we get  
\begin{equation}
\frac{df(t)}{dt}=-\frac{\mu g}{\omega_0A_0}|\sin(\omega_0t+\varphi_0)|-\frac{b}{m}f(t)\sin^2(\omega_0t+\varphi_0)-\frac{D\omega_0A_0}{m}f^2(t)|\sin(\omega_0t+\varphi_0)|^3\,.
\label{firstF1}
\end{equation}
Since, by assumption, the amplitude decreases only slightly over time intervals $\Delta T=T_0=2\pi/\omega_0$, we can make an additional approximation in the equation \eqref{firstF1} by averaging the trigonometric functions over time intervals $\Delta T/2$, as is done in \cite{Wang} for each damping force individually, i.e. 
\begin{equation}
\frac{df(t)}{dt}=-\frac{\mu g}{\omega_0A_0}\langle|\sin(\omega_0t+\varphi_0)|\rangle-\frac{b}{m}f(t)\langle\sin^2(\omega_0t+\varphi_0)\rangle-\frac{D\omega_0A_0}{m}f^2(t)\langle|\sin(\omega_0t+\varphi_0)|^3\rangle\,,
\label{firstF2}
\end{equation}
where we use the notation $\langle A(t)\rangle=(T_0/2)^{-1}\int_t^{t+T_0/2}A(t')dt'$. Since $\langle|\sin(\omega_0t+\varphi_0)|\rangle=2/\pi$, $\langle\sin^2(\omega_0t+\varphi_0)\rangle=1/2$, and $\langle|\sin(\omega_0t+\varphi_0)|^3\rangle=4/(3\pi)$, we get   
%
%
%
\begin{equation}
\frac{df(t)}{dt}=-\left(c_2f^2(t)+c_1f(t)+c_0\right)\,,
\label{firstF3}
\end{equation}
where
\begin{equation}
c_0=\frac{2\mu g}{\pi\omega_0A_0}\,,\,c_1=\frac{b}{2m}\,,\,c_2=\frac{4D\omega_0A_0}{3\pi m}\,,
\label{coeffC}
\end{equation}
as the final differential equation for the function $f(t)$. Since the function $f(t)$ monotonically decreases and the maximal value is $f(0)=1$, the weak damping condition $|df(t)/dt|\ll\omega_0$ $\forall t$ can be written as
\begin{equation}
c_0+c_1+c_2\ll\omega_0\,.
\label{wcondition}
\end{equation}
The condition \eqref{wcondition} depends on the initial conditions due to $c_0\propto A_0^{-1}$ and $c_2\propto A_0$, i.e. in case of a damping force with terms non-linear in velocity, initial conditions play an important role in the overall dynamics of the system. The differential equation \eqref{firstF3} can be solved by separation of variables and integration, i.e. we have to solve 
\begin{equation}
\int_{f(0)}^{f(t)}\frac{df(t')}{c_2f^2(t')+c_1f(t')+c_0}=-\int_0^tdt'\,.
\label{firstF4}
\end{equation}

We note here that an approximate solution of the form \eqref{Xansatz1} passes through the equilibrium position at the same instants as the corresponding undamped solution \eqref{xHO}, but it generally has turning points (i.e. maxima and minima) at earlier time instants compared to turning points of the undamped solution. This fact can be understood by qualitative consideration of the function \eqref{Xansatz1}, its derivative \eqref{Vansatz1}, and using the fact that $df(t)/dt<0$ for all $t\geq0$. The function \eqref{Xansatz1} describes oscillations that decay within the envelope given by $\pm A_0f(t)$. Let us assume that $A_0>0$ and that the function \eqref{Xansatz1} has a turning point, e.g. a local maximum, at instant $t_{max}$, i.e. its derivative \eqref{Vansatz1} is equal to zero at $t_{max}$. The maximum $x(t_{max})$ must be under the envelope, i.e. $x(t_{max})<A_0f(t_{max})$ must hold, because if it were on the envelope, since the envelope has some finite negative derivative at $t_{max}$ and function \eqref{Xansatz1} has derivative equal to zero at that instant by assumption, immediately after $t_{max}$, i.e. for a slightly later instant, the function \eqref{Xansatz1} would cross over the envelope due to the smaller magnitude of its derivative compared to the derivative of the envelope. Thus, the function \eqref{Xansatz1} is under the envelope at $t_{max}$ and it will touch the envelope at some slightly later instant $t_n>t_{max}$ for which $x(t_n)=A_0f(t_n)<x(t_{max})$ and $\cos(\omega_0t_n+\varphi_0)=1$ hold. We can conclude that $t_n$ is the instant of one of the turning points of the undamped solution \eqref{xHO} and that the corresponding turning point of the function \eqref{Xansatz1} occurs at a slightly earlier instant. A similar argument can be applied to the turning points corresponding to the local minima of the function \eqref{Xansatz1}. Thus, the function \eqref{Xansatz1}, with $f(t)$ given by \eqref{firstF4}, generally (for any values of coefficients \eqref{coeffC}, and for any initial conditions) has turning points at earlier instants than the corresponding undamped solution \eqref{xHO}, but for weak enough damping these differences will be negligible.
%
%
%
%

\section{Combinations of three damping forces}
\label{all3}

We consider the case with $c_0>0$, $c_1>0$, and $c_2>0$. Depending on the magnitude of $C=4c_2c_0-c_1^2$, the solutions of \eqref{firstF4} have three different forms. For $C>0$, we get 
\begin{equation}
f_1(t)=\frac{1}{2c_2}\left[\sqrt{C}\tan\left(-\frac{\sqrt{C}}{2}t+\arctan\left(\frac{2c_2+c_1}{\sqrt{C}}\right)\right)-c_1\right]\,,
\label{rjes1}
\end{equation}
for $C<0$  
\begin{equation}
f_2(t)=\frac{1}{2c_2}\left[\sqrt{-C}\tanh\left(\frac{\sqrt{-C}}{2}t+\text{arctanh}\left(\frac{2c_2+c_1}{\sqrt{-C}}\right)\right)-c_1\right]\,,
\label{rjes2}
\end{equation}
and for $C=0$ ($c_1=2\sqrt{c_2c_0}$)  
\begin{equation}
f_3(t)=\frac{1}{\sqrt{c_2}}\left[\left(\sqrt{c_2}\,t+\frac{1}{\sqrt{c_2}+\sqrt{c_0}}\right)^{-1}-\sqrt{c_0}\right]\,.
\label{rjes3}
\end{equation}
%

It is well known that the dynamics of a harmonic oscillator damped (only) by sliding friction halts in finite time. In this case (i.e., for $\mu>0$, $b=0$, and $D=0$) equation \eqref{HOeq} has exact solutions, but we will not engage in a detailed discussion about where and when this exact solution halts, it has already been thoroughly covered elsewhere; see, e.g., \cite{Lapidus, AviAJP, Coulomb, Grk2}. The dynamics necessarily halts in some finite time also for the combination of sliding friction and the other two damping forces due to the constant contribution of sliding friction to the energy loss rate \eqref{Power}, i.e. the contribution of sliding friction to the energy loss rate does not decrease with decreasing velocity, as is the case with the other two damping forces. Thus, depending on the initial conditions, the system will lose all the initial energy and settle down in the equilibrium position, or it will settle at some finite $x(t_{halt})$, for which $|kx(t_{halt})|\leq\mu mg$ and $v(t_{halt})=0$ hold, with elastic potential energy $kx^2(t_{halt})/2$ left in the system. If we randomly choose the initial conditions, the second scenario is of course more likely, because the first scenario can only occur for specially tailored initial conditions. In this short qualitative discussion and throughout the paper, we take, for simplicity, that the dynamic and static friction coefficients are the same, i.e. both are equal to $\mu$.

The functions \eqref{rjes1}, \eqref{rjes2}, and \eqref{rjes3} monotonically decrease from value $f_i(0)=1$ to value $f_i(\tau_i)=0$, where 
\begin{equation}
\tau_1=\frac{2}{\sqrt{C}}\,\arctan\left(\frac{2c_2\sqrt{C}}{C+c_1(c_1+2c_2)}\right)\,,
\label{tau1}
\end{equation}
\begin{equation}
\tau_2=\frac{2}{\sqrt{-C}}\,\text{arctanh}\left(\frac{2c_2\sqrt{-C}}{C+c_1(c_1+2c_2)}\right)\,,
\label{tau2}
\end{equation}
\begin{equation}
\tau_3=\frac{1}{\sqrt{c_0}\left(\sqrt{c_0}+\sqrt{c_2}\right)}\,.
\label{tau3}
\end{equation}
For $t>\tau_i$, the functions $f_i(t)$ become increasingly negative, i.e. the energy \eqref{Energy1} increases in magnitude for $t>\tau_i$ as if we are supplying energy to the system, which is not physical. Therefore, we can take
\begin{equation}
x_i(t)=A_0f_i(t)\theta_i(t)\cos(\omega_0t+\varphi_0)\,
\label{rjesenje123}
\end{equation}
%
%
%
and
\begin{equation}
E_i(t)=\frac{m\omega_0^2A^2_0f^2_i(t)\theta_i(t)}{2}\,,
\label{energija123}
\end{equation}
where
\begin{equation} \label{theta}
    \theta_i(t) = \begin{cases}
\begin{tabular}{@{}cl@{}}
   $1$\, & if\, $0\leq t \leq\tau_i$ \\
    $0$\, & if\, $t>\tau_i$ \, 
\end{tabular}
    \end{cases}
    \end{equation}
and $i=\lbrace 1,2,3\rbrace$, as final expressions for displacements and energies corresponding to functions \eqref{rjes1}, \eqref{rjes2}, and \eqref{rjes3}. 
\begin{figure}[h!t!]
\begin{center}
\includegraphics[width=0.48\textwidth]{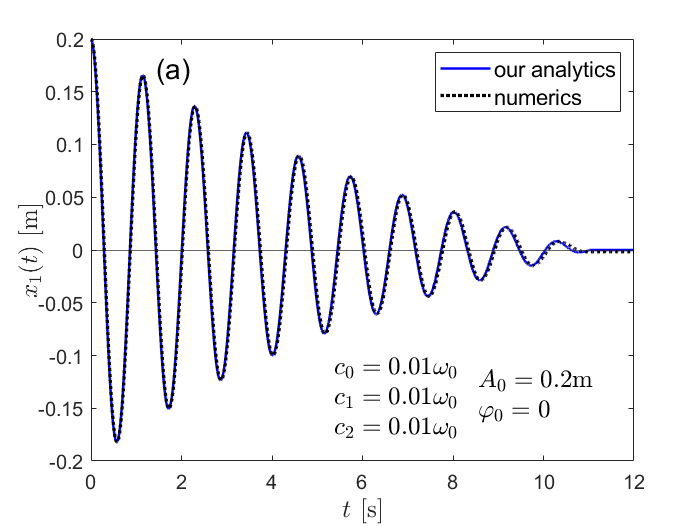}
\includegraphics[width=0.48\textwidth]{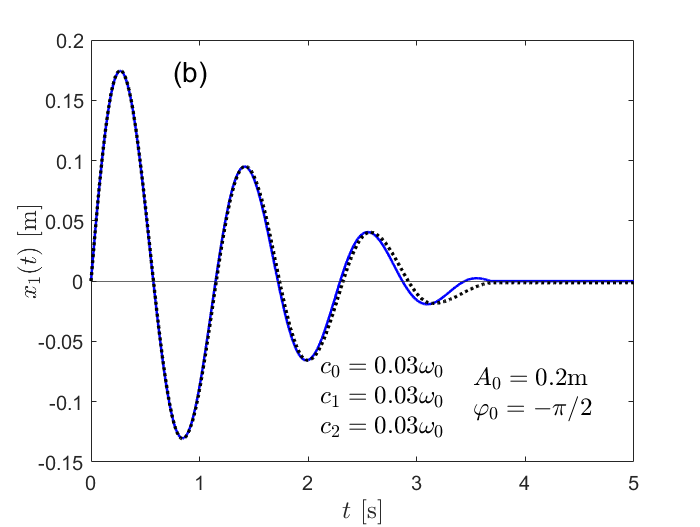}
\includegraphics[width=0.48\textwidth]{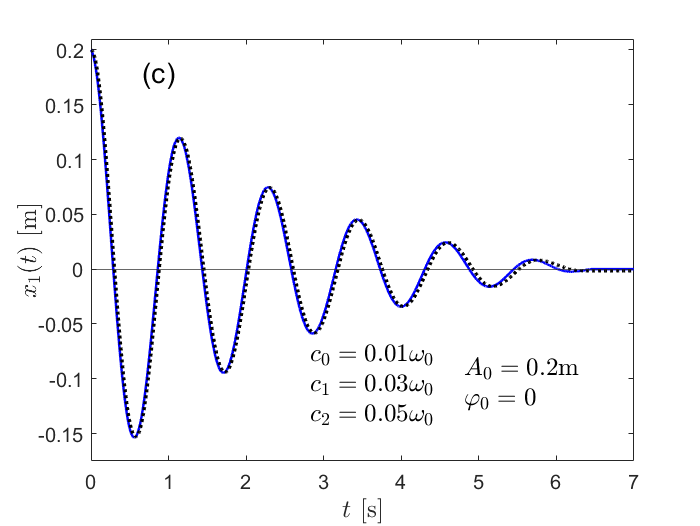}
\includegraphics[width=0.48\textwidth]{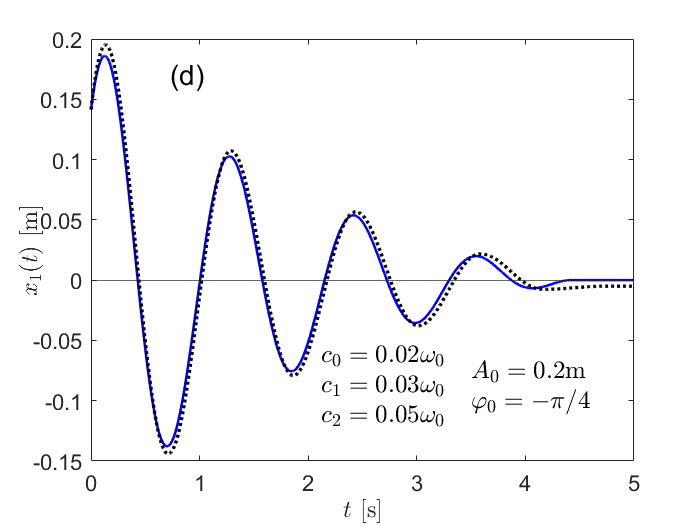}
\end{center}
\caption{Solid blue curves show the solution \eqref{rjesenje123} in the case $C=4c_2c_0-c_1^2>0$, i.e. $x_1(t)$, with $A_0=0.2\,m$ and $\varphi_0=\lbrace 0,-\pi/4,-\pi/2\rbrace$, for various choices of $c_0$, $c_1$, and $c_2$. The dotted black curves in all figures show the numerical solutions of the equation \eqref{HOeq} for the same initial conditions and with the damping constants $\mu$, $b$, and $D$ that correspond to the chosen values of $c_0$, $c_1$, and $c_2$. See text for details.} 
\label{slika1}
\end{figure}
Thus, in our approximate description of the dynamics of a harmonic oscillator damped by three damping forces, the oscillations halt in the equilibrium position, regardless of the initial conditions. This is a shortcoming of our approach because, as we commented, the exact dynamics, given by the equation \eqref{HOeq}, does not halt at the equilibrium position in general. However, as we will show, for sliding friction that is weak enough, this discrepancy is negligible. 

As an example, we consider a block of mass $m=1\,kg$ attached to a spring of stiffness $k=30\,N/m$, and we use $g=9.81\,m/s^2$. The corresponding undamped angular frequency is $\omega_0=5.48\,s^{-1}$ and the period of the undamped system is $T_0=1.15\,s$. These values are along the lines of experiments performed on block-spring systems, e.g. see \cite{Kamela}. The coefficients $c_0$ and $c_2$ depend on $A_0$. Therefore, we take $A_0=0.2\,m$, which corresponds to the initial energy $E_0=m\omega_0^2A_0^2/2=0.6\,J$, and we keep this value fixed in what follows. In Fig.\,\ref{slika1}(a)-(d) the solid blue curves show our approximate analytical solution \eqref{rjesenje123} for the case $i=1$, i.e. the displacement $x_1(t)$, for various values of coefficients $c_n$ and initial phases $\varphi_0$. In Fig.\,\ref{slika1}(a) we consider initial conditions with $\varphi_0=0$, i.e. $\left(x_0=A_0\,,\,v_0=0\right)$, and $c_n=0.01\omega_0$ $\forall n=\lbrace 0,1,2\rbrace$. Using relations \eqref{coeffC}, we determine that the choice $c_0=c_1=c_2=0.01\omega_0$ corresponds to the damping constants $\mu=0.01$, $b=0.11\,kg\,s^{-1}$, and $D=0.12\,kg\,m^{-1}$ (the numbers are rounded to two decimals). In Fig.\,\ref{slika1}(b)-(d) we show the behavior of $x_1(t)$ for other choices of coefficients $c_n$ and for initial conditions with $\varphi_0=-\pi/2$, i.e. for $\left(x_0=0\,,\,v_0=\omega_0A_0\right)$, and $\varphi_0=-\pi/4$, i.e. for $\left(x_0=A_0\sqrt{2}/2\,,\,v_0=\omega_0A_0\sqrt{2}/2\right)$. The dotted black curves in Fig.\,\ref{slika1}(a)-(d) show the numerical solutions of the equation \eqref{HOeq} for the same initial conditions and with damping constants $\mu$, $b$, and $D$ that correspond to the chosen values of $c_0$, $c_1$, and $c_2$. The \emph{ode45} MATLAB function has been utilized for the numerical part of our analysis. \emph{ode45} is a versatile tool for solving ordinary differential equations (ODEs) in general, particularly non-stiff problems like a damped harmonic oscillator in a weak damping regime. \emph{ode45} implements the Runge-Kutta (4,5) method, also known as the Dormand-Prince method. It is appropriate for our problem since it works well when the solution does not change too rapidly over short time scales, which is the case here.
\begin{figure}[h!t!]
\begin{center}
\includegraphics[width=0.48\textwidth]{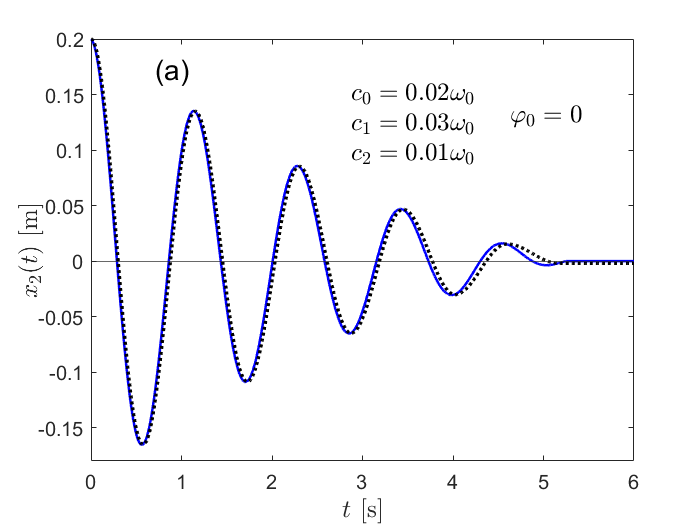}
\includegraphics[width=0.48\textwidth]{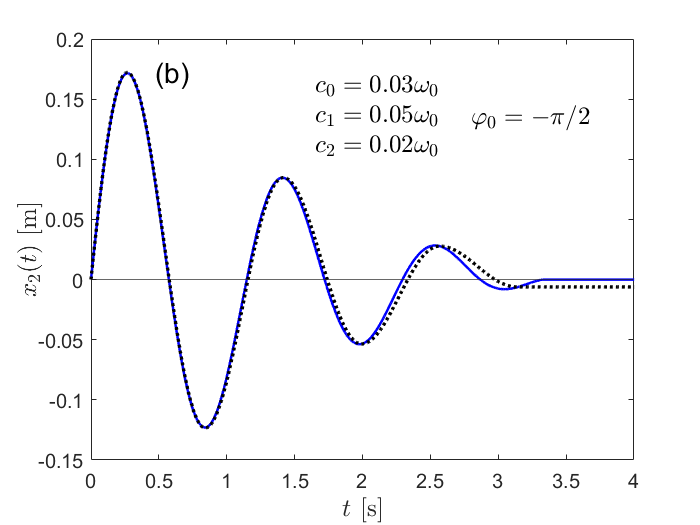}
\end{center}
\caption{Solid blue curves show the solution \eqref{rjesenje123} in the case $C=4c_2c_0-c_1^2<0$, i.e. $x_2(t)$, with $A_0=0.2\,m$ and $\varphi_0=\lbrace 0,-\pi/2\rbrace$, for two choices of $c_0$, $c_1$, and $c_2$. The dotted black curves show the corresponding numerical solutions.} 
\label{slika2}
\end{figure}

\begin{figure}[h!t!]
\begin{center}
\includegraphics[width=0.48\textwidth]{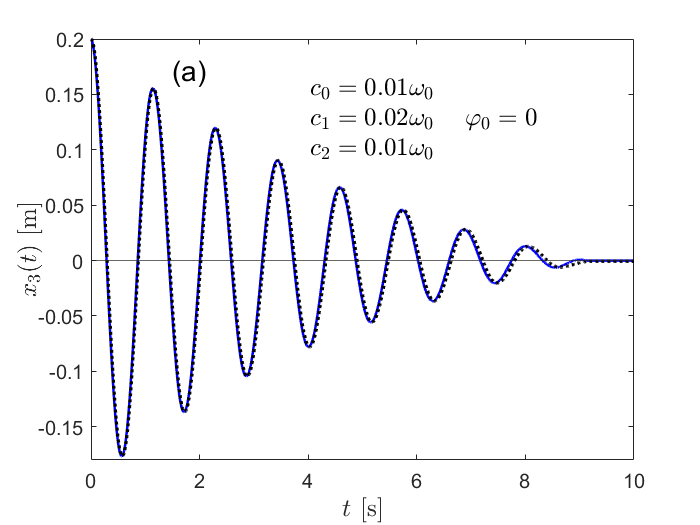}
\includegraphics[width=0.48\textwidth]{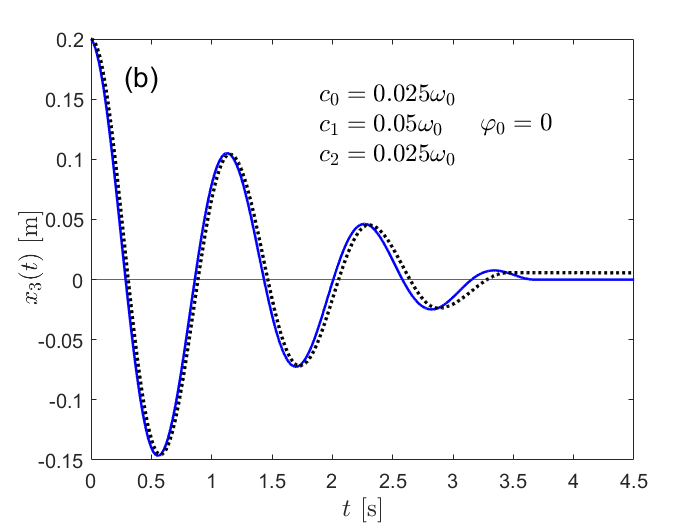}
\end{center}
\caption{Solid blue curves show the solution \eqref{rjesenje123} in the case $C=4c_2c_0-c_1^2=0$, i.e. $x_3(t)$, with $A_0=0.2\,m$ and $\varphi_0=0$, for two choices of $c_0$, $c_1$, and $c_2$. The dotted black curves show the corresponding numerical solutions.} 
\label{slika3}
\end{figure}

In Fig.\,\ref{slika2}(a) and (b) the solid blue curves show our approximate analytical solution \eqref{rjesenje123} for the case $i=2$, i.e. the displacement $x_2(t)$, for two choices of coefficients $c_n$ and initial phases $\varphi_0=\lbrace 0,-\pi/2\rbrace$. In Fig.\,\ref{slika3}(a) and (b) the solid blue curves show our approximate analytical solution \eqref{rjesenje123} for the case $i=3$, i.e. the displacement $x_3(t)$, for two choices of coefficients $c_n$ and initial phase $\varphi_0=0$. In both Fig.\,\ref{slika2} and Fig.\,\ref{slika3}, the dotted black curves show the corresponding numerical solutions. In Fig.\,\ref{slika1}-\,\ref{slika3}, we can see that our approximate analytical solutions give an excellent estimate of the duration of the motion, i.e. of the instant when the dynamics halts. The error of our approximate analytical solutions in the halting position compared to the corresponding numerical solutions can be at most $\Delta x=\pm\mu mg/k$, which amounts, e.g., to $\Delta x=\pm1.6\%\,A_0$ for $\mu=0.01$ and the values of $m$ and $k$ that we use here, i.e. for $c_0=0.01\omega_0$. 
%

%
%
%
%

\begin{figure}[h!t!]
\begin{center}
\includegraphics[width=0.48\textwidth]{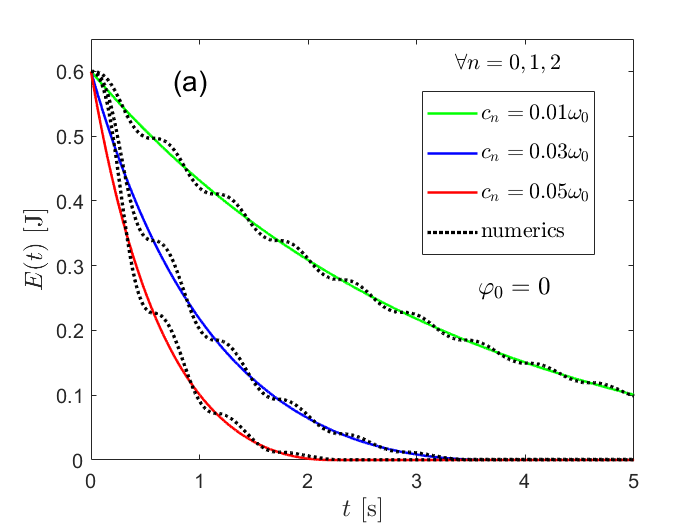}
\includegraphics[width=0.48\textwidth]{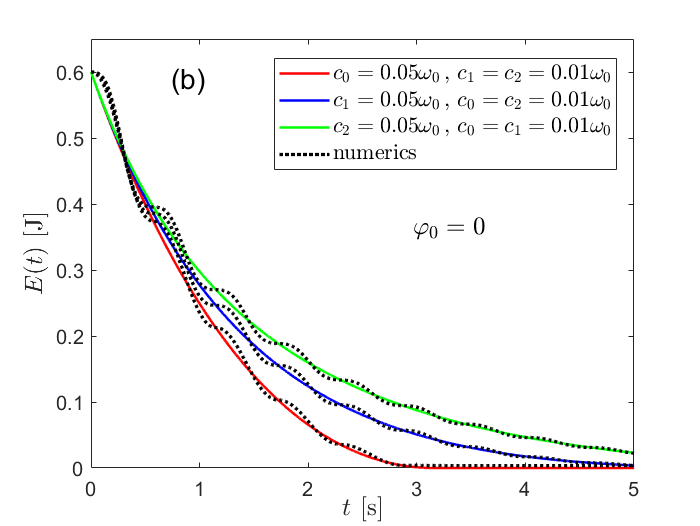}
\end{center}
\caption{Solid colored curves show the energy \eqref{energija123}, with $A_0=0.2\,m$ and $\varphi_0=0$, for different choices of $c_0$, $c_1$, and $c_2$. The black dotted curves show the corresponding numerically obtained energy. (a) All coefficients $c_n$ are equal, i.e. $c_0=c_1=c_2$. (b) Combinations of $c_n$ with one dominant coefficient, i.e. one coefficient has five times higher magnitude than the other two. See text for details.} 
\label{slika4}
\end{figure}

In Fig.\,\ref{slika4} solid colored curves show our approximate energies \eqref{energija123}, and black dotted curves show the corresponding numerically obtained energies. In Fig.\,\ref{slika4}(a) the chosen coefficients $c_n$ satisfy $c_0=c_1=c_2$, thus all solid colored curves in Fig.\,\ref{slika4}(a) correspond to energies $E_1(t)$ (i.e. \eqref{energija123} with $i=1$) calculated for three different values of coefficients $c_n$. In Fig.\,\ref{slika4}(b) the solid red curve corresponds to $E_1(t)$ with $c_0=0.05\omega_0$ and $c_1=c_2=0.01\omega_0$, the solid blue curve corresponds to $E_2(t)$ (i.e. energy \eqref{energija123} with $i=2$) with $c_1=0.05\omega_0$ and $c_0=c_2=0.01\omega_0$, and the solid green curve corresponds to $E_1(t)$ with $c_2=0.05\omega_0$ and $c_0=c_1=0.01\omega_0$. We see in Fig.\,\ref{slika4}(b) that the energy dissipation is the fastest in the case with dominant sliding friction. 
%

\section{Combinations of two damping forces}
\label{comb2}

If $c_1=0$, $c_0>0$, and $c_2>0$, i.e. for a combination of constant and quadratic damping, the solution of \eqref{firstF4} is   
\begin{equation}
f_4(t)=\sqrt{\frac{c_0}{c_2}}\tan\left(-\sqrt{c_0c_2}\,t+\arctan\sqrt{\frac{c_2}{c_0}}\right)\,,
\label{rjes4}
\end{equation}
and in this case our approximate description of the dynamics halts at 
\begin{equation}
\tau_4=\frac{1}{\sqrt{c_0c_2}}\arctan\left(\sqrt{\frac{c_2}{c_0}}\right)\,.
\label{tau4}
\end{equation}
If $c_2=0$, $c_0>0$, and $c_1>0$, i.e. for a combination of constant and linear damping, the solution of \eqref{firstF4} is   
\begin{equation}
f_5(t)=\frac{1}{c_1}\left[\left(c_0+c_1\right)e^{-c_1 t}-c_0\right]\,,
\label{rjes5}
\end{equation}
and the dynamics halts at 
\begin{equation}
\tau_5=\frac{1}{c_1}\ln\left(1+\frac{c_1}{c_0}\right)\,.
\label{tau5}
\end{equation}
Thus, the functions \eqref{rjes4} and \eqref{rjes5} behave qualitatively the same as \eqref{rjes1}, \eqref{rjes2}, and \eqref{rjes3}, i.e. approximate solutions $x_i(t)=A_0f_i(t)\theta_i(t)\cos(\omega_0t+\varphi_0)$, where $i=\lbrace 4,5\rbrace$, settle in equilibrium position at $t=\tau_i$ due to sliding friction. 

\begin{figure}[h!t!]
\begin{center}
\includegraphics[width=0.47\textwidth]{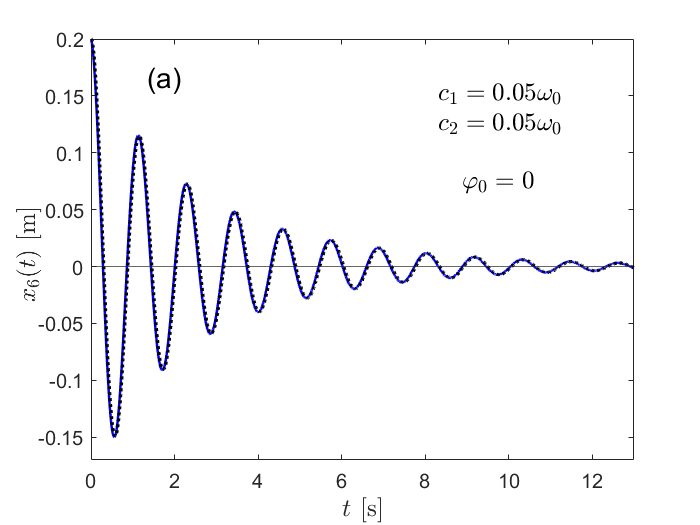}
\includegraphics[width=0.47\textwidth]{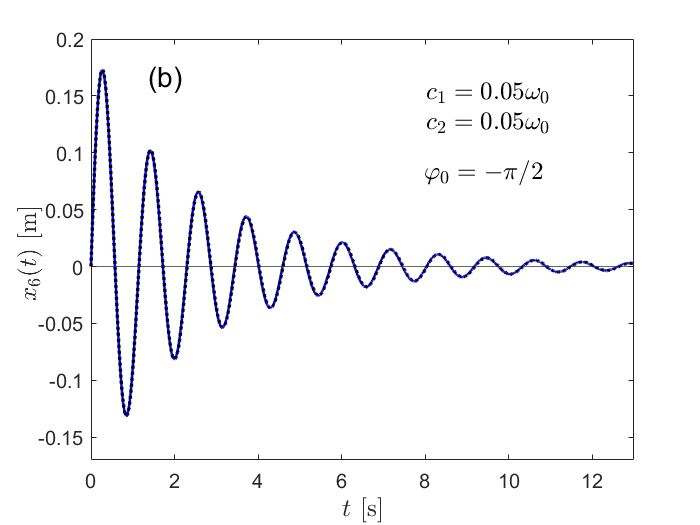}
\includegraphics[width=0.47\textwidth]{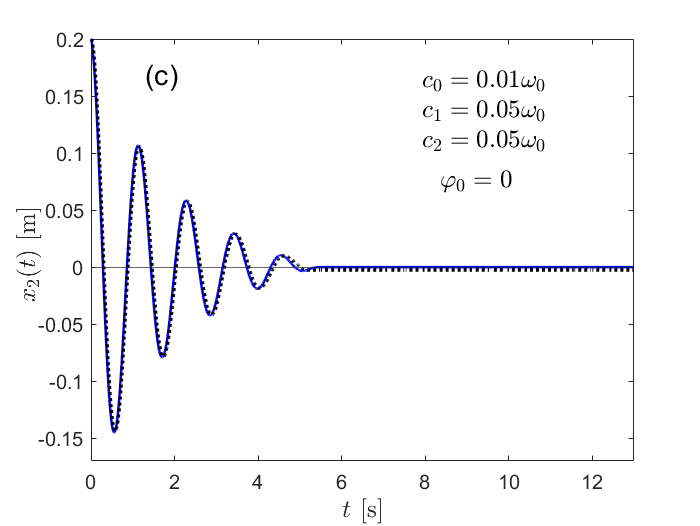}
\includegraphics[width=0.47\textwidth]{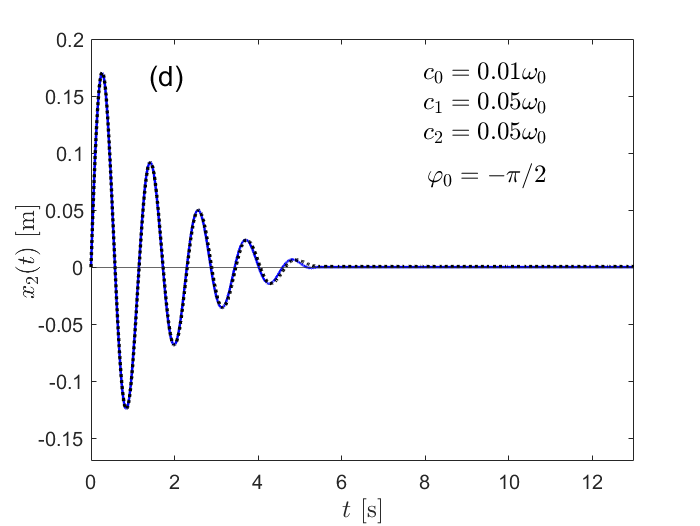}
\end{center}
\caption{Solid blue curves in (a) and (b) show the solution \eqref{X6}, i.e. $x_6(t)$, with $A_0=0.2\,m$ and $\varphi_0=\lbrace 0, -\pi/2\rbrace$, for $c_1=c_2=0.05\omega_0$. Solid blue curves in (c) and (d) show the solution \eqref{rjesenje123} with $A_0=0.2\,m$ and $\varphi_0=\lbrace 0, -\pi/2\rbrace$, for $c_0=0.01\omega_0$, and $c_1=c_2=0.05\omega_0$ (in this case $C=4c_2c_0-c_1^2<0$, i.e. we take \eqref{rjesenje123} with $i=2$). The dotted black curves show the corresponding numerical solutions. Analytical and numerical solutions almost completely overlap. See text for details.} 
\label{slika5}
\end{figure}

In case $c_0=0$, $c_1>0$, and $c_2>0$, i.e. for a combination of linear and quadratic damping, the solution of \eqref{firstF4} is   
\begin{equation}
f_6(t)=\frac{c_1}{e^{c_1 t}(c_1+c_2)-c_2}\,.
\label{rjes6}
\end{equation}
We can easily see that the function \eqref{rjes6} asymptotically approaches zero, thus the corresponding approximate solution
\begin{equation}
x_6(t)=A_0f_6(t)\cos(\omega_0t+\varphi_0)\,
\label{X6}
\end{equation}
asymptotically approaches to equilibrium position. In Fig.\,\ref{slika5}(a) and (b) the solid blue curves show the solution \eqref{X6} with $A_0=0.2\,m$ and $\varphi_0=\lbrace 0, -\pi/2\rbrace$, for $c_1=c_2=0.05\omega_0$. We can see excellent agreement of analytical solution \eqref{X6} and corresponding numerical solutions, i.e. the solid blue and dotted black curves overlap completely. In Fig.\,\ref{slika5}(c) and (d), we see that the addition of sliding friction significantly changes the overall dynamics, although the coefficient $c_0$ has a value five times lower than the value of $c_1$ and $c_2$. 

\section{Limitations of the presented approach and a comment on possible improvements}
\label{dodatak}

The presented approach is based on the assumption that the oscillation frequency and the initial phase in the weakly damped regime remain approximately the same as in the undamped case, while the amplitude gradually decreases, i.e. it is based on the assumption that a function of the form \eqref{Xansatz1} can be taken as a good approximation for the solution of equation \eqref{HOeq} in the case of weak damping. The results shown in the figures presented so far justify this assumption, i.e. the agreement between our approximate analytical solutions and numerical ones is better the weaker the total damping is (i.e. the smaller the sum $c_0+c_1+c_2$ is). So far we have not commented on the upper limit values of $c_0$, $c_1$, and $c_2$ for which our approximate solutions provide a good quantitative description of the dynamics of a harmonic oscillator damped simultaneously by all three damping forces. In this section, by considering the properties of solutions of equation \eqref{HOeq} for each damping force individually, we provide an estimate on the limits of applicability of our approximate analytical solutions \eqref{rjesenje123}. 
Undergraduate students are typically familiar with the viscously damped harmonic oscillator, so we first consider the effect of viscous damping on the frequency and initial phase of the harmonic oscillations.
%
%

In case of viscous damping (i.e. for $b>0$, $\mu=D=0$), equation \eqref{HOeq} has exact analytical solutions. We are interested here only in the underdamped, i.e. oscillatory, regime for which $c_1=b/(2m)<\omega_0$ holds. The corresponding exact solution for general initial conditions is given by (e.g. see \cite{Lelas2023})
\begin{equation}
x_{ex}(t)=\sqrt{x_0^2+\frac{(v_0+c_1x_0)^2}{\omega^2}}e^{-c_1t}\cos(\omega t+\varphi)\,,
\label{xud}
\end{equation}
where
\begin{equation}
\omega=\omega_0\sqrt{1-\left(\frac{c_1}{\omega_0}\right)^2}
\label{omegaV}
\end{equation}
is the damped angular frequency, and
\begin{equation}
\varphi=\arctan\left(-\frac{v_0+c_1x_0}{\omega x_0}\right)
\label{fazaV}
\end{equation}
is the initial phase. Thus, viscous damping reduces the frequency of oscillations compared to undamped oscillations, i.e. $\omega<\omega_0$, by an amount that depends only on $c_1$, while the initial phase $\varphi$ changes in relation to the initial phase of undamped oscillations $\varphi_0$ depending on both the initial conditions and the value of $c_1$. We can easily see that $\omega$ decreases slowly with increasing $c_1$, e.g. $\omega=0.995\omega_0$ for $c_1=0.1\omega_0$. The differences in the initial phases $\varphi$ and $\varphi_0$ in case $c_1>0$ are more subtle than the differences in $\omega$ and $\omega_0$. For example, for initial conditions $(x_0=0,v_0>0)$, we have $\varphi=\varphi_0=-\pi/2$ for any $c_1>0$, while for initial conditions $(x_0>0,v_0=0)$ and, e.g., $c_1=0.1\omega_0$ we have $\varphi=\arctan(-c_1/\omega)=-0.1$ and $\varphi_0=0$. Thus, the function $\arctan(-c_1/\omega)$ decreases faster than $\omega$ with increasing $c_1$. Therefore, in addition to changing the frequency, depending on the initial conditions, viscous damping can significantly change the initial phase, and both effects must be taken into account when estimating the range of values of $c_1$ for which \eqref{Xansatz1} is a valid approximate form of the solution of equation \eqref{HOeq}. This issue has recently been addressed in \cite{LelasPezer} where an approximate solution of the same form as \eqref{Xansatz1} was used to model harmonic oscillations damped (only) with viscous damping, and it was shown that for $0<c_1\lesssim0.1\omega_0$ the modeled solutions excellently describe the exact solution \eqref{xud} with initial conditions $(x_0>0,v_0=0)$ and $(x_0=0,v_0>0)$. Thus, based on the analysis presented in \cite{LelasPezer}, we can also take $c_1\approx0.1\omega_0$ as the upper limit value of $c_1$ in our approximate solutions \eqref{rjesenje123}.

In case of sliding friction (i.e. for $\mu>0$, $b=D=0$) and initial conditions $(x_0>0,v_0=0)$, the exact analytical solution of equation \eqref{HOeq} is (e.g. see \cite{Grk2})
\begin{equation}
x_{ex}(t)=\left(x_0-(2n-1)\frac{\mu mg}{k}\right)\cos(\omega_0t)-(-1)^n\frac{\mu mg}{k}\,
\label{xsf}
\end{equation}
where $n\geq1$ is an integer that represents the number of half periods, i.e. for $0\leq t\leq T_0/2$ we take $n=1$, for $T_0/2\leq t\leq T_0$ we take $n=2$, etc., where $T_0=2\pi/\omega_0$. We have already commented on one shortcoming of our approach with respect to sliding friction, i.e. our approximate solutions \eqref{rjesenje123} describe oscillations that halt at equilibrium position, while the exact dynamics with sliding friction generally does not halt at equilibrium position. Here we will comment on some other shortcomings of our solutions \eqref{rjesenje123} with respect to sliding friction by analyzing solution \eqref{xsf}. It is easy to show that the solution \eqref{xsf} has turning points (maxima and minima) at the instants $t_n=n\pi/\omega_0=nT_0/2$, i.e. at the same instants as the corresponding undamped solution $x_{HO}(t)=x_0\cos(\omega_0t)$. Thus, in terms of the time interval between successive turning points, the sliding friction does not change the frequency of the oscillations compared to the undamped oscillations. What is slightly harder to see is that the solution \eqref{xsf} passes through the equilibrium position somewhat later than the corresponding undamped solution and that this time difference increases in each subsequent half period. The corresponding undamped solution passes through the equilibrium position at instants $\tilde{t}_n=(2n-1)\pi/(2\omega_0)$. If we put $t=\tilde{t}_n+\Delta t_n$ in \eqref{xsf} and equate it to zero, we get
\begin{equation}
\Delta t_n=\frac{1}{\omega_0}\arcsin\left(\frac{1}{\frac{kx_0}{\mu mg}-(2n-1)}\right)\,.
\label{dtsf}
\end{equation}
In case of weak damping, i.e. for $kx_0/(\mu mg)\gg1$, in general we have some small value $\Delta t_1>0$ for $n=1$, and the values of $\Delta t_n$ increase in magnitude with increasing $n$. The argument of the the arcsine function in \eqref{dtsf} must be less than or equal to one, so the maximum integer $n_{max}$ that can appear in \eqref{dtsf} is the largest integer less than $kx_0/(2\mu mg)$. Thus, if the undamped solution passes through the equilibrium position at instants $\tilde{t}_n$, solution \eqref{xsf} will pass it at instants $\tilde{t}_n+\Delta t_n$. These "time shifts" increase with $n$, and the largest is obtained for $n_{max}$. Accordingly, solution \eqref{xsf} passes through the equilibrium position $n_{max}$ times and, sometime after the last passage through the equilibrium position, halts at some finite displacement. As we commented in the last paragraph of Section \ref{Basic}, approximate solutions of the form \eqref{Xansatz1}, unlike the solution \eqref{xsf}, have turning points somewhat earlier than the undamped solution and pass through the equilibrium position at the same instants as the undamped solution. Given these differences and the aforementioned error of our approximate analytical solutions in the halting position, the sliding friction, i.e. the value of $c_0$, significantly affects the quantitative precision of our approximate description of the dynamics governed by equation \eqref{HOeq}. Still, in Fig.\,\ref{slika1}-\,\ref{slika3} we can see that the quantitative agreement between our approximate analytical solutions and numerical solutions is fairly good for values $c_0\lesssim0.03\omega_0$, and in Fig.\,\ref{slika4} we see that our approximate energies \eqref{energija123} follow the behavior of the numerically obtained energies well even for $c_0=0.05\omega_0$. Furthermore, for $c_0\approx0.03\omega_0$ (which corresponds to $\mu\approx0.03$ for our choice of $m$, $k$, and $A_0$) the error of our solutions in the halting position can be at most $\Delta x\approx\pm5\%\,A_0$. If we consider this value of $\Delta x$ to be acceptable, we can take $0<c_0\lesssim0.03\omega_0$ to be the range of values of $c_0$ for which our solutions \eqref{rjesenje123} describe quantitatively well the influence of sliding friction.  
\begin{figure}[h!t!]
\begin{center}
\includegraphics[width=0.55\textwidth]{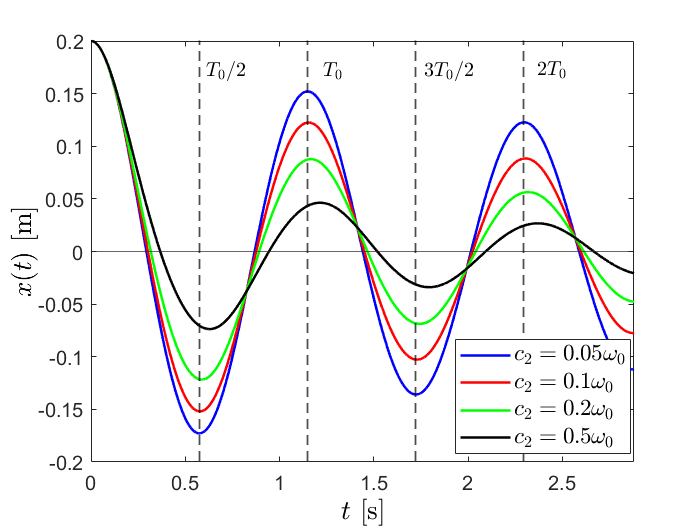}
\end{center}
\caption{Numerical solutions of equation \eqref{HOeq} with initial conditions $(x_0=A_0,v_0=0)$ for $c_2/\omega_0=\lbrace0.05, 0.1, 0.2, 0.5\rbrace$ and $c_0=c_1=0$. Vertical dashed lines are positioned at half periods of the corresponding undamped harmonic oscillator. See text for details.} 
\label{kvadratno1}
\end{figure}

In case of quadratic damping (i.e. for $D>0$, $\mu=b=0$), equation \eqref{HOeq} is a nonlinear differential equation that cannot be solved analytically. The influence of quadratic damping on the frequency of harmonic oscillations can be studied by approximate analytical methods and by considering the time intervals between two consecutive turning points, i.e. half-cycles, but such an approach is mathematically demanding, e.g. see \cite{Smith, AJPNelson}. Here, for simplicity, we will use the numerical solutions of equation \eqref{HOeq} to gain insight into the influence of quadratic damping on harmonic oscillations and estimate the upper limit value of $c_2$ for which our approach gives a good quantitative description of the dynamics. In Fig.\,\ref{kvadratno1} we show numerical solutions of equation \eqref{HOeq} with initial conditions $(x_0=A_0,v_0=0)$ for $c_2/\omega_0=\lbrace0.05, 0.1, 0.2, 0.5\rbrace$ and $c_0=c_1=0$. We can see that for values $c_2/\omega_0=\lbrace0.05, 0.1\rbrace$ the frequency of numerical solutions remains practically the same as in the undamped case, i.e. blue and red curves in Fig.\,\ref{kvadratno1} have all turning points positioned on the dashed vertical lines that denote instants of half periods of the undamped harmonic oscillator. Furthermore, we can notice that the numerical solution for $c_2=0.2\omega_0$, i.e. green curve, has a slight shift of the turning points to the right, while for $c_2=0.5\omega_0$, i.e. black curve, we can clearly see that the influence of quadratic damping is most pronounced during the first half cycle and that the time shift of the first turning point is "transferred" to the subsequent turning points, i.e. after the first half cycle the numerical solution for $c_2=0.5\omega_0$ continues to oscillate with a frequency that is roughly the same as frequency of the undamped oscillator. Based on this brief analysis of the numerical solutions shown in Fig.\,\ref{kvadratno1}, we can safely take $0<c_2\lesssim0.1\omega_0$ to be the range of values of $c_2$ for which our solutions \eqref{rjesenje123} provide a good quantitative description of the influence of quadratic damping on harmonic oscillations. 
\begin{figure}[h!t!]
\begin{center}
\includegraphics[width=0.48\textwidth]{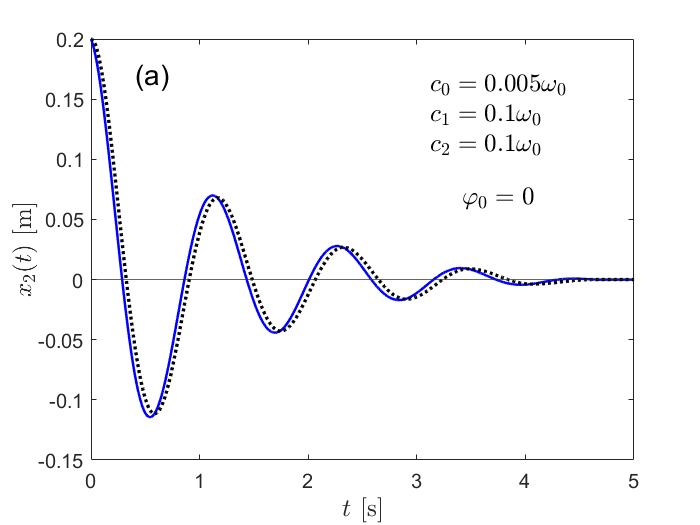}
\includegraphics[width=0.48\textwidth]{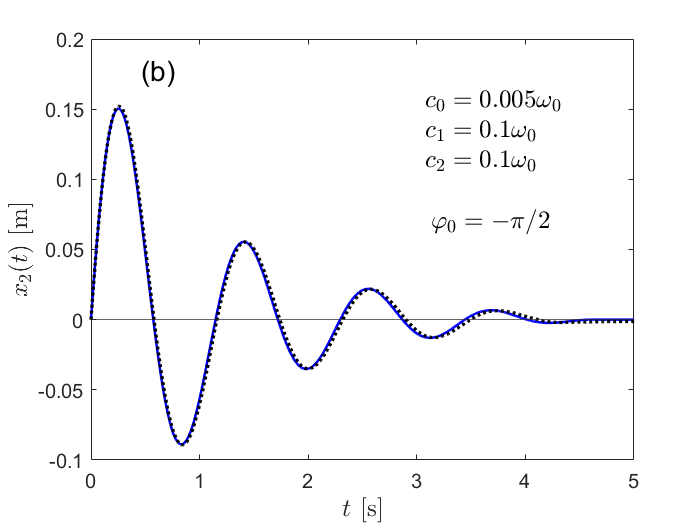}
\includegraphics[width=0.48\textwidth]{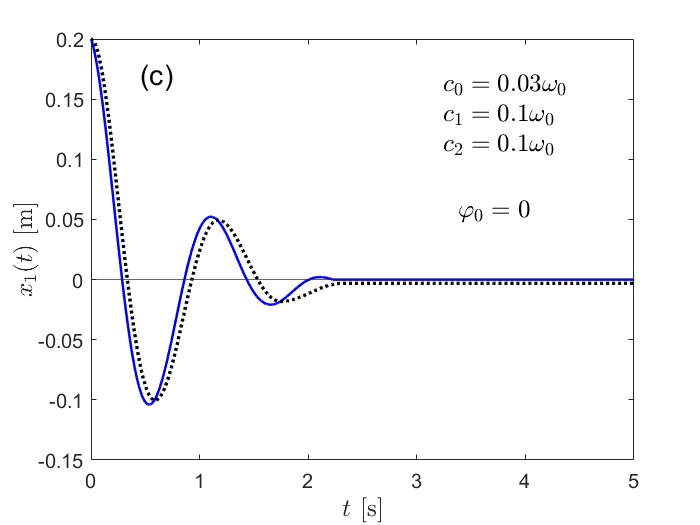}
\includegraphics[width=0.48\textwidth]{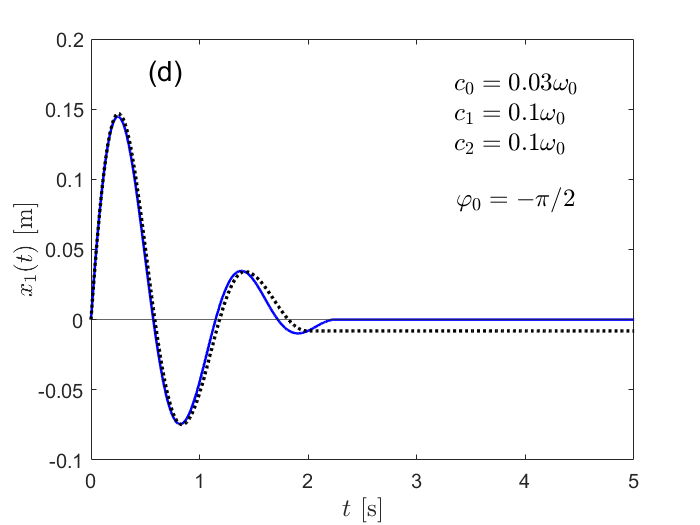}
\end{center}
\caption{Solid blue curves show the solutions \eqref{rjesenje123} with $A_0=0.2\,m$, $c_1=0.1\omega_0$ and $c_2=0.1\omega_0$ in all figures, while $c_0=0.05\omega_0$ in (a) and (b), and $c_0=0.03\omega_0$ in (c) and (d). Furthermore, $\varphi_0=0$, i.e. solutions with initial conditions $(x_0=A_0,v_0=0)$ are shown in (a) and (c), while $\varphi_0=-\pi/2$, i.e. solutions with initial conditions $(x_0=0,v_0=\omega_0A_0)$ are shown in (b) and (d). The dotted black curves show the corresponding numerical solutions.} 
\label{zadnja}
\end{figure}

In Fig.\,\ref{zadnja} we show results that suggest that we gave a reliable estimate of the upper bounds of the coefficients $c_0$, $c_1$, and $c_2$, i.e. our approximate solutions \eqref{rjesenje123} provide a good quantitative description of the dynamics for $0<c_0\lesssim0.03\omega_0$, $0<c_1\lesssim0.1\omega_0$, and $0<c_2\lesssim0.1\omega_0$. Once again, by comparing Fig.\,\ref{zadnja}(a) and (b) with Fig.\,\ref{zadnja}(c) and (d), we can see a strong influence of sliding friction on the precision of our approximate solutions, but even for our chosen upper limit value $c_0=0.03\omega_0$ in combination with upper limit values $c_1=c_2=0.1\omega_0$, i.e. for quite a high value $c_0+c_1+c_2=0.23\omega_0$, our approximate solutions describe the overall dynamics well and give a good estimate of the duration of motion.

Finally, we comment on possible improvements to the presented approach. As we have already mentioned, adding the viscous damping to the harmonic oscillator can affect the initial phase more than the frequency of the oscillations. Thus, the effect of viscous damping alone already suggests that more accurate approximate solutions could be obtained if, instead of a approximate solution of the form \eqref{Xansatz1}, we use
\begin{equation}
x(t)=Af(t)\cos(\omega_0t+\Phi)\,,
\label{Xnew}
\end{equation}
where $A\neq A_0$ and $\Phi\neq\varphi_0$, i.e. if we take into account that the initial phase changes, relative to $\varphi_0$, due to weak damping, and we keep the terms of first order in $|df(t)/dt|\omega_0^{-1}$ in velocity and energy, i.e. through the derivation of the function $f(t)$ by the time averaging of the energy dissipation rate. In such an approach, in addition to the initial conditions, $A$ and $\Phi$ will also depend on the damping constants $\mu$, $b$, and $D$, i.e. on the coefficients $c_0$, $c_1$, and $c_2$. Overall, these are nontrivial corrections to the approach used in this paper and are worth investigating on their own. The examples shown in Fig.\,\ref{zadnja}(a) and (c) clearly indicate that taking into account the change in the initial phase could result in a better agreement between approximate analytical solutions and numerical solutions.

\section{Discussion and conclusion}
\label{Conclusion}

By comparing our analytical solutions with numerical solutions, we have shown that our analytical solutions give an excellent description of the dynamics of a harmonic oscillator simultaneously damped by sliding friction and damping forces that are linear and quadratic in velocity if $c_0+c_1+c_2\ll\omega_0$ is fulfilled. As a paradigmatic example, we considered the block-spring system with damping force \eqref{Fd}. An example of an experiment with the block-spring system damped (only) by sliding friction is given in \cite{Kamela}, where oscillations of a system consisting of two carts and a wooden block on a track were considered. In such an experiment, sliding friction corresponding to small values of $c_0$ (e.g. to $c_0\sim0.01\omega_0$) could be easily realized by taking a lighter wooden block or completely removing it and leaving only the cart to oscillate, since the damping of the cart itself corresponds to weak sliding friction \cite{Kamela}. By adding, for example, small sails to such a system, to increase air resistance, a combination of all three damping forces can be realized, and the values of the coefficients $c_1$ and $c_2$ can be influenced by choosing the shape and size of the sail \cite{Wang}, as well as the initial conditions, since $c_2$ depends on $A_0$. Our results can, of course, be easily applied to all other physical systems described by the damped harmonic oscillator equation. For example, in \cite{Wang} all three types of damping were individually experimentally investigated with physical pendulums. The harmonic oscillator damped by constant damping was realized by a physical pendulum with sliding friction in the support, oscillations with linear damping were realized by magnetically damped physical pendulum (damping by eddy currents), and oscillations of a physical pendulum with (dominantly) quadratic damping were achieved with air resistance \cite{Wang}. The values of the physical parameters and initial conditions used in \cite{Wang} correspond to $c_0\ll\omega_0$, $c_1\ll\omega_0$, and $c_2\ll\omega_0$. Therefore, using a physical pendulum, with a combination of the mentioned effects, it is possible to design a system that corresponds to a harmonic oscillator simultaneously damped by three damping forces and with $c_0+c_1+c_2\ll\omega_0$ fulfilled. Thus, our theoretical results can be tested through experiments suitable for students at the undergraduate level. 

In conclusion, we showed how to obtain an excellent analytical approximation of the amplitude decay, i.e. of the solutions that describe the dynamics of harmonic oscillator simultaneously damped with linear and nonlinear damping forces. Our results are suitable for undergraduate physics courses, since we use basic mathematical concepts familiar to students at this level. The presented approach is a good example of the importance of modeling in physics, i.e. it can be emphasized to students that even when strong assumptions are used, models can be obtained that give a very good quantitative description of complex phenomena. Based on our theoretical results, laboratory exercises can be designed in which an interesting task for students could be the measurement of the duration of free oscillations and the analysis of the validity of approximate analytical expressions \eqref{tau1}-\eqref{tau3} depending on the parameters of the system and initial conditions. Furthermore, vibrations simultaneously damped with linear and nonlinear damping forces are a topic of wider interest that goes beyond physics teaching. For example, a paper \cite{JSVCoulombViscous} was recently published in which oscillations with a combination of sliding friction and linear damping were considered due to their importance in the context of engineering applications and the amplitude decay of the same form as the function \eqref{rjes5} was derived. Since the approach we use is mathematically simpler than the phase space coordinate averaging used in \cite{JSVCoulombViscous}, and also easily enables the addition of a third damping force, i.e. quadratic damping, we believe that it can be useful in the context of applications in engineering as well. Finally, in the last paragraph of Section \ref{dodatak}, we commented on possible improvements to the approach presented in this paper, which will be the topic of our future work.

\section{Acknowledgments}

This work was supported by the QuantiXLie Center of Excellence, a project co-financed by the Croatian Government and European Union through the European Regional Development Fund, the Competitiveness and Cohesion Operational Programme (Grant No. KK.01.1.1.01.0004).




\bibliographystyle{unsrt}
\bibliography{PublicationList.bib}

\begin{thebibliography}{10}

\bibitem{Cutnell8}
J.D. Cutnell and K.W. Johnson.
\newblock {\em Physics}.
\newblock John Wiley \& Sons, 2009.

\bibitem{Resnick10}
David Halliday, Robert Resnick, and Jearl Walker.
\newblock {\em Fundamentals of Physics}.
\newblock John Wiley \& Sons, 2013.

\bibitem{Young2020university}
Hugh~D. Young and Roger~A. Freedman.
\newblock {\em University Physics with Modern Physics}.
\newblock Pearson, 2020.

\bibitem{Waves}
Frank~S. Crawford.
\newblock {\em Waves: Berkeley Physics Course. volume 3}.
\newblock McGraw-Hill, New York, 1968.

\bibitem{Lapidus}
I.~R. Lapidus.
\newblock Motion of a harmonic oscillator with sliding friction.
\newblock {\em American Journal of Physics}, 38(11):1360–1361, nov 1970.

\bibitem{AviAJP}
A.~Marchewka, David.~S. Abbott, and R.~J. BeichnerKamela.
\newblock Oscillator damped by a constant-magnitude friction force.
\newblock {\em American Journal of Physics}, 72(4):477–483, apr 2004.

\bibitem{Grk2}
A.~Anastasios~Adamopoulosa and N.~Adamopoulos.
\newblock Constant and quadratic damping of free oscillations: easy solutions.
\newblock {\em International Journal of Mathematical Education in Science and Technology}, 53(11):3151--3161, jan 2022.

\bibitem{Kamela}
M.~Kamela.
\newblock An oscillating system with sliding friction.
\newblock {\em The Physics Teacher}, 45(2):110--113, feb 2007.

\bibitem{Smith}
B.~R.~Jr. Smith.
\newblock The quadratically damped oscillator: A case study of a non-linear equation of motion.
\newblock {\em American Journal of Physics}, 80(9), sep 2012.

\bibitem{Mungan}
C.~E. Mungan and T.~C. Lipscombe.
\newblock Oscillations of a quadratically damped pendulum.
\newblock {\em European Journal of Physics}, 34(5):1243–1253, jul 2013.

\bibitem{Wang}
X.~Wang, C.~Schmitt, and M.~Payne.
\newblock Oscillations with three damping effects.
\newblock {\em European Journal of Physics}, 23(2):155--164, jan 2002.

\bibitem{AJPpendulum}
P.~T. Squire.
\newblock Pendulum damping.
\newblock {\em American Journal of Physics}, 54(11):984–991, nov 1986.

\bibitem{Markho}
P.~H. Markho.
\newblock On free vibrations with combined viscous and coulomb damping.
\newblock {\em Journal of Dynamic Systems, Measurement, and Control}, 102(4):283--286, dec 1980.

\bibitem{AJPRicch}
A.~Ricchiuto and A.~Tozzi.
\newblock Motion of a harmonic oscillator with sliding and viscous friction.
\newblock {\em American Journal of Physics}, 50(2):176–179, feb 1982.

\bibitem{AJPHinrich}
P.~F. Hinrichsen and C.~I. Larnder.
\newblock Combined viscous and dry friction damping of oscillatory motion.
\newblock {\em American Journal of Physics}, 86(8):577–584, aug 2018.

\bibitem{AJPNelson}
R.~A. Nelson and M.~G. Olsson.
\newblock The pendulum—rich physics from a simple system.
\newblock {\em American Journal of Physics}, 54(2):112–121, feb 1986.

\bibitem{Bacon}
M.~E. Bacon and D.~D. Nguyen.
\newblock Real-world damping of a physical pendulum.
\newblock {\em European Journal of Physics}, 26(4):651–655, may 2005.

\bibitem{Vitorino}
M.~V. Vitorino, A.~Vieira, and M.~S. Rodrigues.
\newblock Effect of sliding friction in harmonic oscillators.
\newblock {\em Scientific Reports}, 7:3726, jun 2017.

\bibitem{Coulomb}
J.~A. Rizcallah.
\newblock Revisiting the coulomb-damped harmonic oscillator.
\newblock {\em European Journal of Physics}, 40(5), aug 2019.

\bibitem{Lelas2023}
Karlo Lelas, Nikola Poljak, and Dario Jukić.
\newblock {Damped harmonic oscillator revisited: The fastest route to equilibrium}.
\newblock {\em American Journal of Physics}, 91(10):767--775, 10 2023.

\bibitem{LelasPezer}
K.~Lelas and R.~Pezer.
\newblock Modeling the amplitude and energy decay of a weakly damped harmonic oscillator using the energy dissipation rate and a simple trick.
\newblock {\em European Journal of Physics}, 46(1):015004, dec 2024.

\bibitem{JSVCoulombViscous}
B.~K. Karthik, R.~B. Shreesha, V.~Shrikanth, and A.~K. Gaonkar.
\newblock Prediction of energy dissipation by analytical solution to combined viscous and coulomb damping.
\newblock {\em Journal of Sound and Vibration}, 573:118216, 2024.

\end{thebibliography}

\end{document}